\documentstyle[preprint,aps]{revtex}
\begin{document}
 
\title{Lambda flow in heavy-ion collisions: the role
of final-state interactions}
\bigskip
\author{G. Q. Li and C. M. Ko}
\address{Cyclotron Institute and Physics Department,\\
Texas A\&M University, College Station, Texas 77843, USA}
 
\maketitle
 
\begin{abstract}
Lambda flow in Ni+Ni collisions at SIS energies is studied in 
the relativistic transport model (RVUU 1.0). It is found that 
for primordial lambdas the flow is considerably weaker than proton 
flow. The inclusion of final-state interactions, especially the 
propagation of lambdas in mean-field potential, brings the
lambda flow close to that of protons. An accurate determination of 
lambda flow in heavy-ion experiments is shown to be very useful for 
studying lambda properties in dense matter.
\end{abstract}
 
 
\section{Introduction}

Collective flows of hadrons and light fragments in heavy-ion collisions
have been observed unambiguously at various incident energies
\cite{krof92,gut89,eos1,eos2,fopi,ags,dan86}.  Detailed comparisons
of theoretical predictions with the experimental data have already provided
valuable information about the nuclear equation of state \cite{pan,zhang}, 
the in-medium nucleon-nucleon cross sections \cite{xu},
and the medium modification of hadron properties \cite{li95a}.
It has been found that proton and fragment flows are largely caused by
the compressional pressure generated at high densities
and thus carry the information about the nuclear equation of state.
Proton flow at intermediate energies has also been found to be
sensitive to medium modifications of the nucleon-nucleon cross sections.
Transport model calculations without including mean-filed potentials
have shown that pion flow \cite{bass,liba,dan95}, as well as antikaon
and antiproton flow \cite{rqmd,arc}, are mostly determined by their large
absorption cross sections by nucleons, and thus usually show 
`antiflow' with respect to nucleons. These results are, however,
qualitatively changed when their mean-field
potentials are included in the transport model.
Both the antiflows of pions \cite{fae96} and antikaons \cite{li96}
are found to be significantly reduced by their attractive potentials
in nuclear medium. A similar effect is expected for the
antiflow of antiprotons when their attractive mean-field potential
is included in the study. Also, kaon flow has been shown to be
sensitive to the kaon potential in nuclear medium \cite{li95a}.
A comparison with recent experimental
data from the FOPI collaboration at GSI has shown that
a kaon feels both an attractive scalar and a repulsive vector potential
\cite{brown96}, consistent with the prediction of effective chiral
Lagrangians \cite{rho,waas}.
The resulting kaon potential is weakly repulsive and thus
makes the kaon flow extremely weak compared to nucleon flow.

More recently, collective flow of lambdas has been
measured through the reconstruction of $p\pi^-$ pairs
by the FOPI collaboration in Ni+Ni collisions at 1.93 GeV/nucleon
\cite{fopi}, and by the EOS collaboration in Ni+Cu collisions 
at 2.0 GeV/nucleon \cite{eos2}.
Although the magnitude of lambda flow differs in the two measurements,
both collaborations have found that lambdas
flow in the same direction as nucleons, and its strength is similar to
that of proton flow. The quantitative difference between
the two measurements is
due to different acceptance cuts and centrality selections
in the two experiments.

In heavy-ion collisions around 1-2 GeV/nucleon, lambdas are mainly
produced in association with kaons from baryon-baryon and
meson-baryon collisions.
In Refs. \cite{li95a,brown96} we have studied kaon flow from Ni+Ni
collisions at similar energies, and found that the flow of primordial
kaons is considerably weaker than proton flow. This is
also true for primordial lambdas, as we shall show later.
In the case of kaons, the inclusion of final-state interactions, mainly
their propagation in mean field, tends to repel them away from nucleons,
and this reduces kaon flow or even changes its
sign if the kaon potential is very repulsive.

The final-state interactions of a lambda are different from that of
a kaon. At low energies, the lambda-nucleon cross section
is relatively large compared to the kaon-nucleon cross section.
Also, instead of a weak repulsive potential as for a kaon,
the lambda potential is attractive
in the density and momentum range relevant for
heavy-ion collisions at beam energies around 1-2 GeV/nucleon.
These differences are expected to make
the final lambda flow quite
different from the kaon flow, although they are very similar
without final-state interactions.  In particular, we shall show that
the attractive potential between a lambda and the nuclear
matter makes its flow similar to the nucleon flow.

The purpose of this paper is to study quantitatively the effects
of final-state interactions on lambda flow in heavy-ion
collisions at SIS energies. We also study the sensitivity of
lambda flow to its properties in dense matter, which
are important for understanding the
properties of `strange' stars and the possible kaon condensation
\cite{star1,star2,star3}. In Section II, we briefly review the
relativistic transport model (RVUU 1.0) that has been used extensively
in studying
heavy-ion collisions at SIS energies. We also discuss how we treat
lambda production, scattering, and propagation in the transport model.
The results from this study are presented
in Section III. The paper ends with a short summary in Section IV.

\section{The relativistic transport model}

The relativistic transport model (RVUU 1.0), first developed in
Ref. \cite{ko} and further extended in Refs. \cite{fang94,li94},
is based on the non-linear $\sigma$-$\omega$ model and includes
explicitly the nucleon, delta resonance, and pion. It can also treat
the production of eta, kaon, antikaon, hyperon, antiproton, and dilepton
in heavy-ion collisions at SIS energies using the perturbative test particle
method. 

At SIS energies, lambdas are mainly produced from baryon-baryon
($BB\rightarrow B\Lambda K$, with $B$ denotes either a nucleon
or a delta resonance) and pion-nucleon ($\pi N \rightarrow \Lambda K$)
collisions. The process $\pi\Delta\rightarrow \Lambda K$ is neglected
since its cross section is much smaller than that of $\pi N\rightarrow
\Lambda K$, as shown in the resonance model of Ref. \cite{fae}.
For the isospin-averaged cross sections $\sigma _{BB\rightarrow
B\Lambda K}$, we use the parameterization of Ref. \cite{rand},
e.g.,
\begin{eqnarray}
\sigma _{NN\rightarrow N\Lambda K} = 0.072 {p_{max}\over m_K} ~~{\rm mb},
\end{eqnarray}
with
\begin{eqnarray}\label{bb}
p_{max}= {1\over 2\sqrt s} \Big[\big(s-(m_N+m_\Lambda +m_K)^2\big)
\big(s-(m_N+m_\Lambda -m_K\big)\Big]^{1/2}.
\end{eqnarray}
The cross section for $\pi N\rightarrow \Lambda K$ is taken from the
parameterization in Ref. \cite{cugnon}, i.e.,
\begin{eqnarray}\label{pin}
\sigma _{\pi N\rightarrow \Lambda K}& =& 4.94 (\sqrt s- \sqrt {s_0}) ~{\rm mb}
, ~\sqrt s- \sqrt {s_0} \le 0.091~ {\rm GeV}\\
 & = & {0.045\over 0.01+(\sqrt s- \sqrt {s_0} )} ~{\rm mb}, ~\sqrt s-\sqrt
{s_0} > 0.091 ~{\rm GeV},
\end{eqnarray} 
with $\sqrt {s_0}= m_\Lambda +m_K$.

In the following calculations, we always include the medium modification
of kaon properties by using
its in-medium (pole) mass $m_K^*$ determined from the mean-field
approximation to the chiral Lagrangian \cite{fang94}, i.e.,
\begin{equation}
m_{K}^*\approx m_K[1-\frac{\Sigma_{KN}}{f^2m_K^2}\rho_S
+\frac{3}{4f^2m_K}\rho_N]^{1/2},
\end{equation}
where $f=93$ MeV is the pion decay constant, and $\Sigma_{KN}\approx
350-450$ MeV is the $KN$ sigma term, which depends on the strangeness
content of a nucleon. The scalar and nuclear densities are denoted,
respectively, by $\rho_S$ and $\rho_N$.

The lambda mean-field potential in nuclear medium has been extracted
from the properties of hypernuclei \cite{dover,hyper}. It has also
been studied in the Dirac-Brueckner approach using a boson-exchange
model for the $\Lambda N$ potential \cite{julich}, or by extending
the Walecka model from SU(2) to SU(3) \cite{rufa,glen}.
In the naive constituent quark model, the lambda mean-field potential
is about 2/3 of that of a nucleon, and is in qualitative agreement with
the empirical observation \cite{dover,hyper}. 
To see the sensitivity of lambda flow to the lambda potential in dense
matter, we adopt the constituent quark relation but
introduce a parameter $\alpha$ in the lambda vector potential, i.e.,
\begin{eqnarray}
\Sigma _S^\Lambda \approx 2/3 \Sigma _S^N, ~~
\Sigma _V^\Lambda \approx 2/3 \alpha \Sigma _V^N, 
\end{eqnarray}
where $\Sigma _S^N$ and $\Sigma _V^N$ are the nucleon scalar and vector
potentials in the non-linear $\sigma$-$\omega$ model \cite{li94}.
The lambda optical-model potential can then be defined as
\begin{eqnarray}
U_\Lambda ({\bf p}, \rho ) =\Big((m_\Lambda -\Sigma _S^\Lambda )^2
+{\bf p}^2\Big)^{1/2}+\Sigma _V^\Lambda 
-\Big(m_\Lambda^2+{\bf p}^2\Big)^{1/2}.
\end{eqnarray}
We consider three values of $\alpha$, i.e., 0.85, 1.0, and 1.2 so that
the lambda potential is varied within a reasonable range. The corresponding
lambda potential is shown in Fig. 1. The nucleon potential used in this
study is based on the parameter set in Ref. \cite{li94} that corresponds
to a soft equation of state with a compression modulus of 200 MeV
and a nucleon effective mass of $0.83m_N$ at normal nuclear matter
density. The solid circle with error bars is the currently determined
lambda potential in nuclear matter from both the structure of hypernuclei
\cite{dover,hyper} and the Dirac-Brueckner calculation \cite{julich}.

Including the medium modification of the lambda, as well as the nucleon
and kaon, $\sqrt s$ in Eq. (\ref{bb}) is replaced by
$\sqrt {s^*} = (m_{N_1}^{*2}+{\bf p}^2)^{1/2}+(m_{N_2}^{*2}+{\bf p}^2)^{1/2}
+(1-2/3\alpha )\Sigma _V^N$, while $m_N$, $m_\Lambda$, and $m_K$
are replaced by $m_N^*=m_N-\Sigma _S^N$, $m_\Lambda ^*= m_\Lambda
-\Sigma _S^\Lambda$, and $m_K^*$, respectively.
Similarly, $\sqrt s$ and $\sqrt {s_0}$ in Eq. (\ref{pin})
are replaced by $\sqrt {s^*} = (m_N^{*2} + {\bf p}^2)^{1/2} 
+(m_\pi^2 + {\bf p}^2)^{1/2}+ (1-2/3\alpha )\Sigma _V^N$ 
and $\sqrt {s_0^*} = m_\Lambda ^* + m_K^*$, respectively.
In calculating lambda production, we have thus included the change
of the threshold due to the medium modification of hadron properties.

After production, a lambda interacts with surrounding
baryons. This includes both $\Lambda N$ scattering and lambda propagation
in the mean field generated by nuclear medium.
In principle, the $\Lambda N$ scattering is modified in medium,
but we have not attempted to include this effect in a consistent fashion.
As in most transport models, the free cross section will be used.
However, we will show results from varying the $\Lambda N$ cross section.
Furthermore, we will not address the question of
consistently separating the interactions into
a mean-field and a scattering part in the transport model simulation,
which has recently been raised in Ref. \cite{pawel}.

The $\Lambda N$ cross section has been measured
experimentally \cite{data} and also studied in
a boson-exchange model \cite{julich}. For
beam energies considered in the present study, the $\Lambda N$ inelastic
scattering, mainly $\Lambda N\rightarrow \Lambda \Delta$, is relatively
unimportant and is neglected. We thus consider only $\Lambda N$ elastic
scattering. The experimental data can be fitted by
\begin{eqnarray}\label{lamn}
\sigma _{\Lambda N}=12.0+{0.43\over p_{lab}^{3.3}} ~~ {\rm mb}, 
\end{eqnarray}
where $p_{lab}$, in units of GeV, is the momentum of the lambda 
in nucleon rest frame. This cross section is shown in Fig. 2, where
circles are data from Ref. \cite{data} and the solid curve is 
the parameterization. The angular distribution for the $\Lambda N$
scattering is assumed to be isotropic in its center-of-mass frame,
which is consistent with the results from
the boson-exchange model of Ref. \cite{julich}.
To see how the lambda flow might depend on the angular distribution
in $\Lambda N$ scattering, we have carried out
a calculation by taking the $\Lambda N$ differential cross section
as the proton-proton one \cite{bert88}, and
we find that this has very little effect on the final lambda flow.

Including a mean-field potential,
the lambda equations of motion between collisions are then given by
\begin{eqnarray}
{d{\bf x}\over dt} = {{\bf p}^*\over E^*}, ~~~ {d{\bf p}\over dt}
= -\nabla _x U_\Lambda ({\bf p}, \rho ),
\end{eqnarray}
where $E^*=\sqrt{{\bf p}^{*2}+(m_\Lambda -\Sigma _S^\Lambda )^2}.$

\section{results and discussions}

Using the model described in the above, we have carried out
a calculation of lambda production in Ni+Ni collisions at 1.93
GeV/nucleon and impact parameter $b\le$ 4 fm that corresponds
approximately to the centrality selection in the FOPI experiment.
A direct comparison of the predicted lambda flow
with the preliminary FOPI and EOS data requires also knowledge on
the experimental acceptance cuts.
Since these corrections apply similarly to protons
and lambdas, we have not attempted to include them in this study.
What is significant in both FOPI and EOS data is the
fact that the observed lambda flow is in the same direction as
proton flow, and both have similar strengths.
We thus compare instead the predicted lambda and proton flows
from the transport model.

The results for lambda flow, i.e., the average transverse momentum
$\langle p_x \rangle$ as a function of rapidity $y$ in the nucleus-nucleus
center-of-mass frame, are shown in Fig. 3. 
The dashed curve is the flow of primordial lambdas.
The dotted-dashed curve gives the flow of lambdas after including
elastic $\Lambda N$ scattering with a cross section given by
Eq. (\ref{lamn}). The lambda flow including both scattering
and lambda propagation in mean field (with $\alpha = 1.0$)
is shown by the solid curve. For comparison, we also show by 
dotted curve the proton flow.

To be more quantitative, we introduce the
flow parameter defined as the slope of the
average transverse momentum at mid-rapidity, i.e.,
\begin{eqnarray}
F={d\langle p_x \rangle \over dy}|_{y=0},
\end{eqnarray}
As in the case of kaons, the flow of primordial lambdas,
which arises mainly from Lorentz boost in the direction
of the baryon-baryon or meson-baryon pairs that produce the antikaon,
is considerably weaker than the proton flow.  Their flow parameter is
about 60 MeV as compared to the proton flow parameter of about 140 MeV.
Including elastic $\Lambda N$ scattering increases the flow parameter
to about 80 MeV. This is due to the effect of thermalization
that increases the magnitude of lambda momentum. 
The propagation of lambdas in mean-field potential
further enhances their flow in the direction of nucleons, and 
the flow parameter is now about 115 MeV, and is close to that of protons. 
Overall, final-state interactions enhance the lambda flow parameter
by about a factor of two.

The importance of final-state interactions can also be seen from
$\langle p_x \rangle _{max}$, which occurs near the projectile and target
rapidities. The magnitude of $\langle p_x \rangle _{max}$ for 
primordial lambdas is about 45 MeV, which increases to about 
60 MeV when $\Lambda N$ scattering is included, and further
increases to about 100 MeV after including also lambda propagation.
Final-state interactions thus increase
$\langle p_x \rangle _{max}$ by about a factor of two as well.

To learn quantitatively the lambda potential from lambda flow requires
not only more accurate experimental data but also better
theoretical understandings on how the flow is
affected by a possible change of $\Lambda N$ cross section in nuclear
medium. For this purpose, we have carried out two calculations,
one with $\sigma _{\Lambda N}$ reduced by 50\% and the other with 
$\sigma _{\Lambda N}$ increased by 50\% ($\alpha$ is always kept 
at 1.0). The results are shown in Fig. 4. It is seen that changing 
$\Lambda N$ cross section by a factor of 3 does not affect 
significantly the lambda flow, with the flow parameter modified by
only about 15 MeV. On the average, each lambda undergoes
about 1.3, 2.5 and 3.6 collisions for
$0.5\sigma _{\Lambda N}$, $\sigma _{\Lambda N}$,
and $1.5\sigma _{\Lambda N}$, respectively. This implies that with
1.3 $\Lambda N$ scattering, lambdas are already thermalized
by nucleons, so further increase of the cross section
does not have significant effects on the lambda momentum distribution.

To see the sensitivity of lambda flow to the lambda potential
we have done two additional calculations with $\alpha =1.2$ and 0.85,
while always using the $\Lambda N$ cross section given by Eq. (\ref{lamn}).
The results for lambda flow are shown in Fig. 5 together with that
using $\alpha=1.0$.
We see that with an increasingly attractive
lambda potential, the lambda flow becomes stronger.
The flow parameter $F$ changes from about 95 MeV for $\alpha$=1.2
to about 135 MeV for $\alpha$=0.85. It is thus possible to
differentiate the scenario with a shallow lambda potential (as given
by $\alpha$=1.2) from that with a deep lambda potential (as given by
$\alpha$=0.85) if the lambda flow can be measured with
good accuracy.

Similarly, one can also look at the azimuthal distribution
of lambdas to examine the effects of lambda potential.
The results for the lambda azimuthal distribution
near the target rapidity are shown in Fig. 6 for three different
potentials, as well as for the case without lambda potential. We have
normalized these results around $\phi =0^0$. 
In all four cases, the lambda azimuthal distribution exhibits
a peak around $\phi =180 ^0$, as in the nucleon distribution
shown in the figure
by the dotted curve. The lambda azimuthal anisotropy gets
more pronounced as the lambda potential becomes more attractive.
To show this more quantitatively, we introduce
an anisotropy parameter $R$ defined by
\begin{eqnarray}
R= {dN/d\phi (\phi =180^0\pm 15^0)\over dN/d\phi (\phi=0^0\pm 15^0)}.
\end{eqnarray}
It is seen that $R$ changes from about 1.9 to about 2.6 when $\alpha$
changes from 1.2 to 0.85.  Without lambda potential $R$ is about 1.4.
For comparison, we note that the anisotropy parameter for nucleons
is $R\approx 3.0$.  Thus, the lambda mean-field potential
also enhances its azimuthal anisotropy in the direction of nucleons.

In Fig. 7, we summarize the dependence of the flow parameter $F$ and
the anisotropy parameter $R$, as well as the lambda yield,
on the strength parameter $\alpha$ of the lambda
vector potential. It is seen that the lambda yield also
depends on the strength of lambda potential. A stronger repulsive
lambda vector potential increases the threshold for its production
and thus decreases its yield. As a result, we see
a similar dependence of the lambda yield on the strength
parameter $\alpha$ as for
the lambda flow parameter and anisotropy anisotropy.

\section{summary}

In summary, we have studied lambda flow in heavy-ion collisions at
SIS energies. We have found that the primordial lambdas show a relatively
weak flow as compared with the nucleon flow. The inclusion
of final-state interactions, especially the propagation in mean-field
potential, enhances the lambda flow in the direction of nucleons,
and brings the theoretical results in agreement with the preliminary data
from both FOPI \cite{fopi} and EOS \cite{eos2} collaboration.
Significant differences in both lambda in-plane and out-of-plane
flows are found between the results with and without lambda potential.
On the other hand, the final lambda flow is relatively insensitive
to changes of the $\Lambda N$ cross section within a reasonable range.
Accurate measurements of both lambda yield and flow
in heavy-ion collisions thus allow us to
determine the lambda potential in the dense matter formed in
heavy-ion collisions.  This information cannot be obtained from
studies of hypernuclei which provide only the lambda potential at and below
normal nuclear matter density.  For understanding the properties of neutron
stars, lambda properties at higher densities than normal nuclear
matter are needed. The information one derives from lambda flow
in heavy-ion collisions is thus very useful for studying
neutron star properties \cite{star1,star2,star3}.

\vskip 1cm
We are grateful to K. Wolf for providing us computational
resources. This work was supported in part by the 
National Science Foundation under Grant No. PHY-9509266.

\newpage
 
\centerline{\bf Figure Captions}
 
{\bf Fig. 1:} The lambda optical potential. The solid circle is the
empirically determined lambda potential at normal nuclear matter
density.
 
{\bf Fig. 2:} The elastic $\Lambda N$ cross section. Circles are experimental
data from Ref. \cite{data}, and the curve is the parameterization
given by Eq. (\ref{lamn}).

{\bf Fig. 3:} The average transverse momentum of lambdas as a function
of rapidity. The dashed, dotted-dashed, and solid curves are for
primordial lambdas, lambdas with scattering, and lambdas
with both scattering and propagation, respectively. The dotted curve
is for protons. 
 
{\bf Fig. 4:} Effects of $\Lambda N$ cross section on lambda
average transverse momentum distribution in rapidity space.
The dotted, solid, and dashed curves are obtained
with $0.5\sigma _{\Lambda N}$, $\sigma _{\Lambda N}$, and 
$1.5\sigma _{\Lambda N}$, respectively.
 
{\bf Fig. 5:} Effects of lambda potential on its
average transverse momentum distribution in rapidity space.
The dotted, solid, and dashed curves are obtained
with $\alpha$=0.85, 1.0 and 1.2, respectively.
 
{\bf Fig. 6:} The azimuthal distribution of lambdas near target rapidity.
The dotted curve gives that of nucleons.

{\bf Fig. 7:} The lambda yield, flow parameter, and azimuthal 
anisotropy parameter as functions of the strength
parameter $\alpha$ for the lambda vector potential.

\end{document}